\documentstyle[12pt,epsfig]{article}

\newcommand{\be}{\begin{eqnarray}}
\newcommand{\ee}{\end{eqnarray}}

\newcommand{\dm}{\mbox{$\Delta m_{21}^2$~}}
\newcommand{\dmsol}{\mbox{$\Delta m^2_{\odot}$~}}

\newcommand{\br}{\mbox{$^{8}{B}~$}}

\newcommand{\kl}{\mbox{KamLAND~}}

\newcommand{\thsol}{\mbox{$\theta_{\odot}$~}}

\newcommand{\sss}{\sin^2 \theta_{\odot}}
\begin{document}

\begin{flushright}
\end{flushright}

\begin{center}
{\Large \bf Solar Neutrino oscillation parameters   
after KamLAND}
\vspace{.5in}

{\bf Srubabati Goswami\footnote{e-mail: sruba@mri.ernet.in},
Abhijit Bandyopadhyay\footnote{e-mail: abhi@theory.saha.ernet.in},
 Sandhya Choubey\footnote{e-mail: sandhya@he.sissa.it}
}
\vskip .5cm

$^1$
{\it Harish-Chandra Research Institute},\\{\it Chhatnag Road, Jhusi,
Allahabad  211 019, INDIA}\\
$^2$
{\it
Theory Group, Saha Institute of Nuclear Physics},\\
{\it 1/AF, Bidhannagar,
Calcutta 700 064, INDIA}\\
$^3${\it INFN, Sezione di Trieste and
Scuola Internazionale Superiore di Studi Avanzati,\\
I-34014,
Trieste, Italy}\\
\vskip 1in

\end{center}

\begin{abstract}
We explore the impact of the 
data from the \kl experiment 
in constraining neutrino mass and mixing 
angles involved in solar neutrino oscillations.  
In particular we discuss the precision 
with which we can determine the  
the mass squared 
difference $\Delta m^2_{\odot}$ and the mixing angle 
$\theta_{\odot}$ from combined solar and \kl data. 
We show that the precision with which $\dmsol$ can be 
determined improves drastically with the \kl data 
but the sensitivity of \kl to the mixing angle 
is not as good. We study the effect of enhanced statistics in \kl 
as well as reduced systematics in improving the precision. 
We also show the effect of the SNO salt data in improving the 
precision. 
Finally we discuss  how a dedicated reactor experiment with a baseline of 70 
km can improve the $\thsol$ sensitivity 
by a large amount.

\end{abstract}

\newpage

\section{Introduction}

Two very important results in the field of neutrino oscillations were 
declared in year 2002. 
The first data from the neutral current(NC) events 
from Sudbury Neutrino Observatory(SNO) experiment were announced in 
April 2002 
\cite{Ahmad:2002jz}
Comparison of the 
the NC event rates with the charged current(CC) event rates  
established the presence of $\nu_\mu/\nu_\tau$ 
component in the solar $\nu_e$ flux reinforcing  the fact that neutrino  
oscillation is responsible for the solar neutrino shortfall 
observed in the Homestake, SAGE, GALLEX/GNO, Kamiokande and SuperKamiokande 
experiments. 
The global analysis of solar neutrino data picked up 
the Large Mixing Angle (LMA) MSW as the 
preferred solution \cite{ncanalysis}. 
The smoking-gun evidence came in December 2002 when the \kl experiment
reported a 
distortion in the reactor anti-neutrino spectrum corresponding to the LMA 
parameters
\cite{Eguchi:2002dm}. 
The induction of the \kl data in the global oscillation analysis 
resulted in splitting  the allowed LMA zone in two parts (at 99\% C.L.)
--
low-LMA lying around  
$\Delta m_\odot^2 = 7.2 \times 10^{-5}~{\rm eV^2}$, 
$\sin^2 \theta_\odot = 0.3$,
and high-LMA with 
$\Delta m_\odot^2 = 1.5 \times 10^{-4}~{\rm eV^2}$,
$\sin^2 \theta_\odot = 0.3$ respectively.
The low-LMA solution was preferred
statistically by the data \cite{Bandyopadhyay:2002en}. 
The recently announced SNO data from the salt-phase \cite{snosalt} 
has further disfavoured high-LMA and it now appears at $>$  99.13\% C.L. 
\cite{snosaltus}. 
Thus the SNO and \kl results have heralded the birth of the precision 
era in the measurement of solar neutrino oscillation parameter. 
In this article we take a closer look at the precision with which
we know the solar neutrino 
oscillation parameters at present and critically examine how precisely they 
can be measured with future data.

\section{Oscillation Parameters from solar neutrino data} 
\begin{figure}[t]
\centerline{\psfig{figure=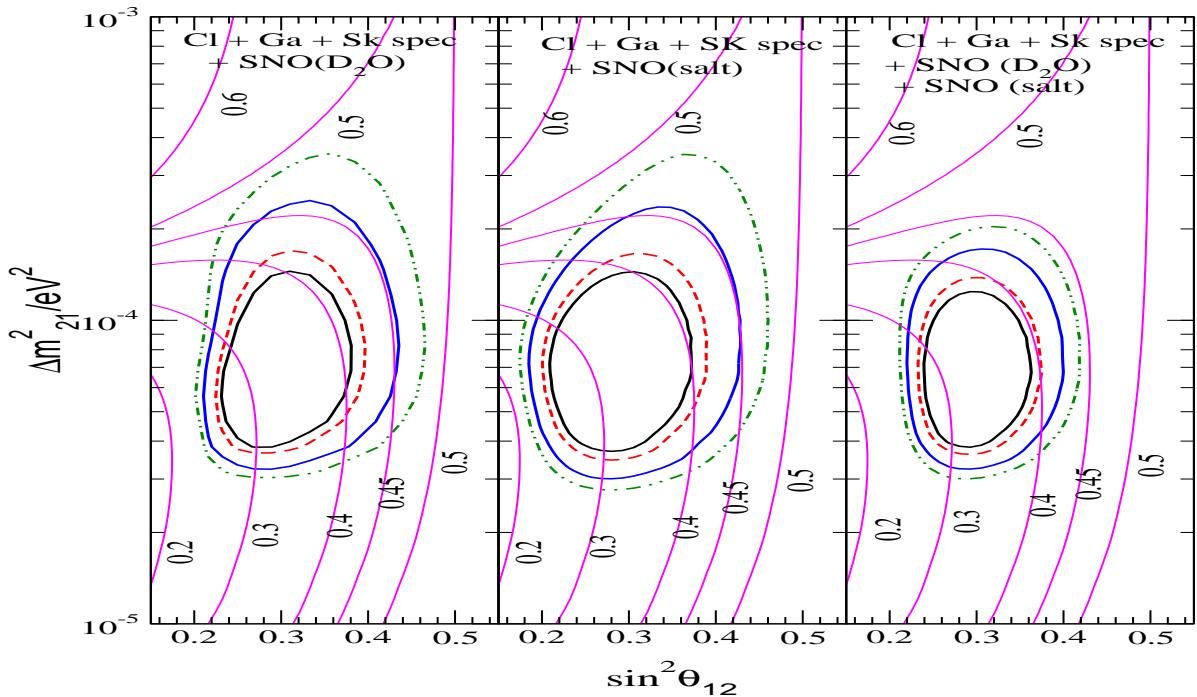,height=5.in,width=7.0in}}
\vskip -0.5cm
\caption{The 90\%, 95\%, 99\% and 99.73\% C.L.
allowed regions in the $\dm-\sin^2\theta_{\odot}$ plane from global
$\chi^2$-analysis of the data from solar neutrino experiments.
We use $\Delta \chi^2$ values to plot the C.L. contours
corresponding to a two parameter fit.
Also shown are the lines of constant CC/NC event rate
ratio $R_{CC/NC}$.  
}
\label{sol2osc}
\end{figure}


In fig. \ref{sol2osc} we show the impact of the SNO NC data from 
the pure $D_2O$ 
phase, the salt phase as well as combining the information from both 
phases on the oscillation parameters $\dmsol(\equiv \dm)$ and $\sss(
\equiv \sin^2\theta_{12}$) from a two-flavour analysis . 
We include the total rates from the radiochemical experiments
Cl and Ga (Gallex, SAGE and GNO combined) \cite{sol} and
the 1496 day 44 bin SK Zenith
angle spectrum data \cite{SKsolar}. 
For the pure $D_2O$ phase we use the CC+ES+NC spectrum data 
whereas for the salt phase we use the published CC,ES and NC rates
\cite{snosaltus}.
The details of the analysis procedure can be found in 
\cite{Choubey:2002nc}. 
Also shown superposed on these curves 
are the isorates of the $CC/NC$ ratio. 
We find that 
\\
$\bullet$ The upper limit on $\Delta m^2_\odot$ tightens 
with the increased statistics when the salt data 
is added to the data from the pure $D_2O$ phase. 
\\
$\bullet$ The upper limit on $\sin^2\theta_{\odot}$ tightens.
For the $^{8}{B}$ neutrinos undergoing adiabatic MSW transition in the 
sun $R_{CC}/R_{NC} \sim \sin^2\theta$. The SNO salt data corresponds to a 
lower value of the $CC/NC$ ratio which results in a shift of $\sss$ 
towards smaller values.

\begin{table}
\begin{center}
\begin{tabular}{c|cc|cc}
\hline\hline
Data & \multicolumn{2}{c|} {best-fit parameters}  & \multicolumn{2} {c}
{99\% C.L. allowed range}\\
\cline{2-5}
set used & \dm/($10^{-5}$eV$^2$) & $\sss$ & \dm/($10^{-5}$eV$^2$) & $\sss$\\
\hline \hline
Cl+Ga+SK+$D_2O$& $6.06$ & 0.29 & $3.2-24.5$ &  $0.21 - 0.44$ \\
Cl+Ga+SK+salt& $6.08$ & 0.28 & $3.0-23.7$ &  $0.19 - 0.43$ \\
Cl+Ga+SK+$D_2O$+salt& $6.06$ & 0.29 & $3.2-17.2$ &  $0.22 - 0.40$ \\
Cl+Ga+SK+$D_2O$+KL& $7.17$ & 0.3 & $5.3-9.9$ &  $0.22 - 0.44$ \\
Cl+Ga+SK+$D_2O$+salt+KL& $7.17$ & 0.3 & $5.3-9.8$ &  $0.22 - 0.40$ \\
\hline \hline
\end{tabular}
\label{tab1}
\caption{
The best-fit values of the solar neutrino oscillation
parameters, obtained using different combinations
of data sets. Shown also are the 99\% C.L.
allowed ranges of the parameters from
the different analyses.}
\end{center}
\end{table}

\section{Impact of  \kl data on oscillation parameters} 

The \kl detector measures the
reactor antineutrino spectrum from Japanese commercial nuclear reactors 
situated at a distance of 80 -800 km. 
In this section we present our results of global two-generation 
$\chi^2$ analysis of solar+\kl spectrum data. 
For details we refer to  our analysis in 
\cite{Bandyopadhyay:2002en,prekl}
. 

\begin{figure}[t]
\centerline{\psfig{figure=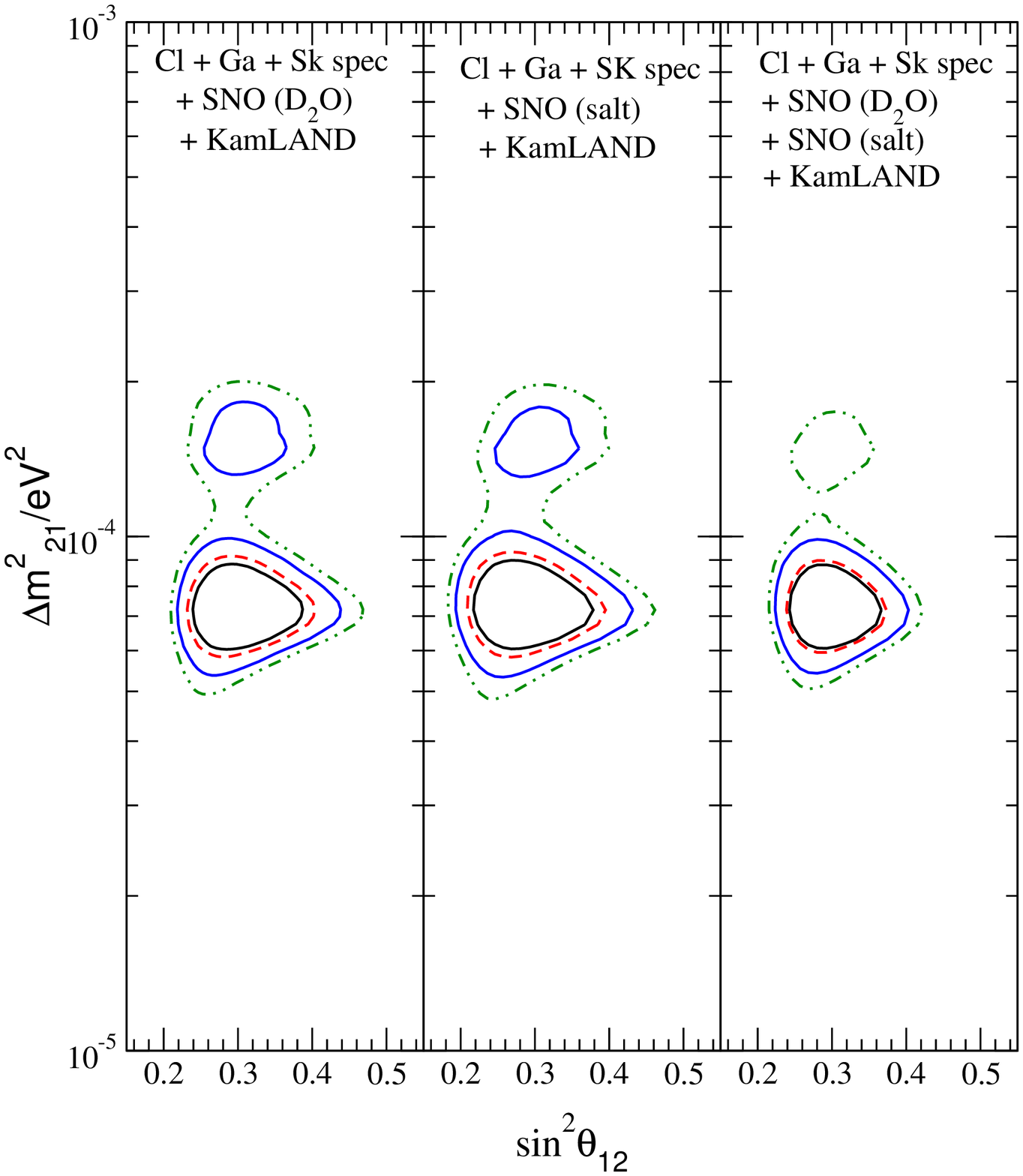,height=5.in,width=7.in}}
\vskip -0.5cm
\caption{The 90\%, 95\%, 99\% and 99.73\% C.L.
allowed regions in the $\dm-\sin^2\theta_{\odot}$ plane from global
$\chi^2$-analysis of solar and \kl data.
We use the $\Delta \chi^2$ values corresponding to a
2 parameter fit
to plot the C.L. contours.  
}
\label{solkl2osc}
\end{figure}

Figure \ref{solkl2osc} shows the allowed regions obtained from global 
solar and  162 Ton-year \kl \\ 
spectrum data.   
As is seen from the leftmost panel of 
figure \ref{solkl2osc}  the inclusion of the \kl data breaks the 
allowed LMA region into two parts at 99\% C.L.. 
The low-LMA region is centered around a best-fit  $\dmsol$ of 
$7.2\times 10^{-5}$ eV$^2$ and the high-LMA region is centered around 
$1.5 \times 10^{-4}$ eV$^2$. At 3$\sigma$ the two regions merge. 
The low-LMA region is statistically preferred over the high-LMA region. 
With the addition of the SNO salt data the high-LMA solution gets disfavoured 
at 99.13\% C.L.. 
In Table \ref{tab1} we show the allowed ranges of $\dmsol$ 
and $\sss$ from solar and combined 
solar+\kl analysis. 
We find that $\Delta m^2_{21}$ is further constrained with the addition of the 
\kl data but $\sin^2\theta_{12}$ is nor constrained any further. 

\section{Closer look at \kl sensitivity}
 
\begin{table}
\begin{center}
\begin{tabular}{ccccc}
\hline
{Data} & 99\% CL &99\% CL  & 99\% CL
& 99\% CL \cr
{set} & range of & spread &range  & spread
\cr
{used} & $\Delta m^2_{21}\times$ & of &
of & in  \cr
& 10$^{-5}$eV$^2$ & {$\Delta m^2_{21}$} & $\sin^2\theta_{12}$
& {$\sin^2\theta_{12}$} \cr
\hline
{only sol} & 3.2 - 17.0
&{68\%} & $0.22-0.40$ &{29\%}\cr
{sol+162 Ty KL}&  5.3 - 9.8
& {30\%}
& $0.22-0.40$ &{29\%}  \cr
{sol+1 kTy KL}& 6.5 - 8.0
& {10\%} &
$0.23-0.39$ & {26\%}\cr
{sol+3 kTy KL}& 6.8 - 7.6
&{6\%} & $0.24-0.37$ & {21\%} \cr
\hline
\end{tabular}
\label{klbounds}
\caption
{The range of parameter values allowed at 99\% C.L.
and the corresponding spread.
}
\end{center}
\end{table}
In Table \ref{klbounds} we take a closer look at the sensitivity of the \kl 
experiment to  the parameters $\Delta m^2_{21}$ and $\theta_{12}$ 
with the current as well as simulated future data and examine how far the 
sensitivity can improve with the future data.
We define the \% spread in oscillation parameters as 
\be
{\rm spread} = \frac{ prm_{max} - prm_{min}}
{prm_{max} + prm_{min}}\times 100
\label{error}
\ee
and determine this quantity for the current solar and \kl data 
as well as increasing the \kl statistics. 
The current systematic error in \kl is 6.42\% and the largest 
contribution comes from the uncertainty in fiducial volume. 
This is expected to improve with the calibration of the fiducial 
volume and we use a 5\% systematic error for 1 kTy simulated \kl data 
and 3\% systematic error for 3 kTy simulated \kl data. 
The table reveals the tremendous sensitivity of \kl to 
$\Delta m^2_{21}$. 
The addition of the present \kl data improves the spread in 
$\Delta m^2_{21}$ to 30\% from 68\% obtained with only solar data.  
With 1 kTy \kl data it improves to 10\% and if we increase the statistics to 
3 kTy then the uncertainty in $\Delta m^2_{21}$ reduces to 6\%. 
However the sensitivity of \kl to the parameter $\theta_{12}$ does not look 
as good. 
The addition of the current \kl data to the global solar analysis 
does not improve the spread in $\sin^2\theta_{12}$. 
With reduction of the systematic error to 5\% the spread with 1 kTy 
statistics improves to 26\% and even with a very optimistic value of 
3\% for the systematic uncertainty and a substantial increase of statistics to 
3 kTy, the \kl data fails to constrain $\theta_{12}$ much better than the current solar neutrino experiments. 

\begin{figure}[p]
\centerline{\psfig{figure=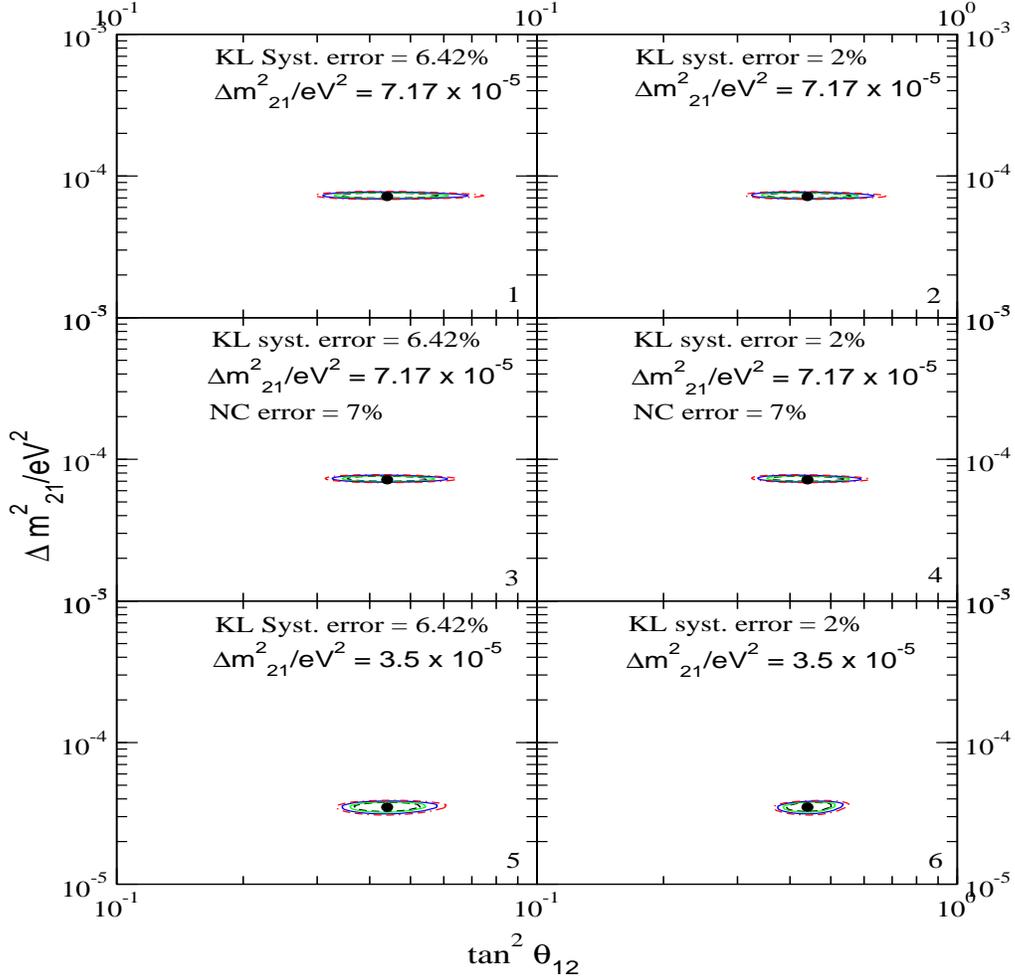,height=6in,width=5.5in}}
\vskip -1cm 
\caption{The contours for
the combined analysis using the solar and 3 kTy
simulated KamLAND spectrum.
The first two rows of panels  correspond to spectrum simulated at 
$7.2\times 10^{-5}$ eV$^2$ 
while the lowermost row of  panels are for \kl data simulated at
$\Delta m_{21}^2=3.5\times 10^{-5}$
eV$^2$. 
}
\label{fict}
\end{figure}

In figure \ref{fict} we compare the allowed areas computed with 
spectrum simulated at $\dm$ = $7.2\times 10^{-5}$ eV$^2$
and $\dm = 3.5\times 10^{-5}$ eV$^2$. 
We show limits for the current \kl
systematic uncertainty of 6.42\% and a very optimistic systematic
uncertainty of just 2\%.
The \% spread in uncertainty for the spectrum simulated at 
$7.2 \times 10^{-5}$ eV$^2$ with 6.42\%
systematic uncertainty is  37\% while for $3.5 \times 10^{-5}$ eV$^2$
case 
the spread is 25\%.
The effect of reducing
the systematics to 2\%  results in the spread coming down to
32\% and 19\% respectively \cite{th12}. 
We would like to mention that the figure \ref{fict} uses   
the CC, NC and ES rates from the $D_2O$ 
and not the latest results from the salt phase. However the purpose of this 
figure is to compare the spread in $\sss$ 
obtained for the two different values of $\dm$ and the use of the 
sno salt phase data is not going to change the relative spreads 
significantly. 
We also present in the middle panel of \ref{fict} the allowed areas drawn 
using a 7\% uncertainty in the NC rate. The uncertainty in the NC rate 
from the $D_2O$(salt) phase data is 12\%(9\%).  

\begin{figure}[t]
\centerline{\epsfig{file=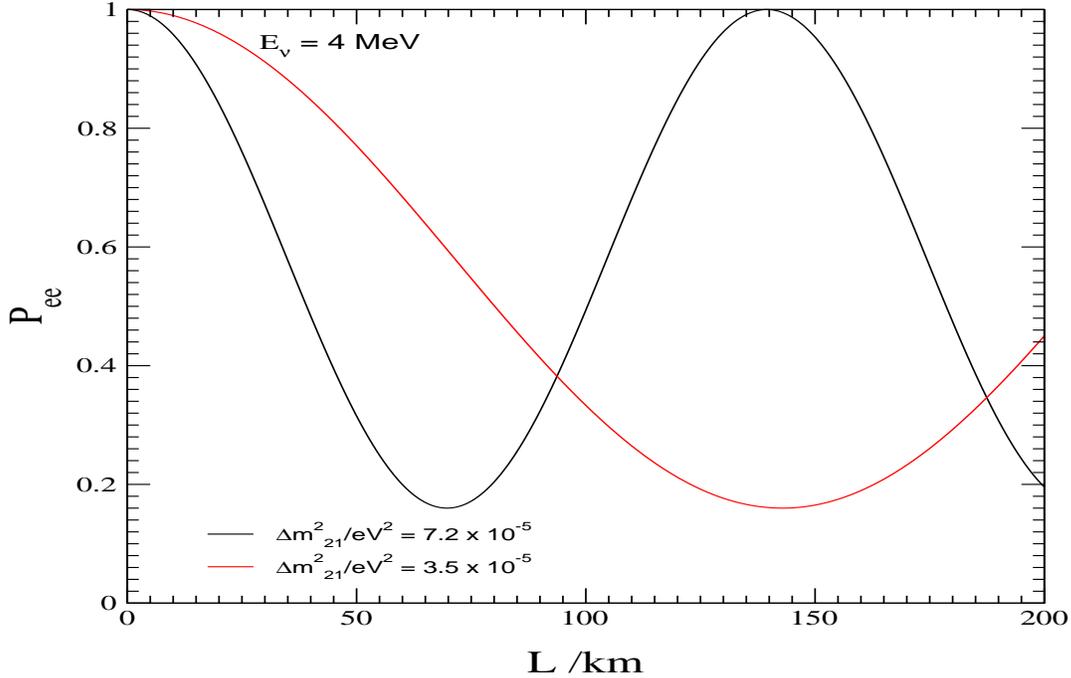,width=6in,height=5.5in}}
\vskip -1cm
\caption{The probability vs distance for an average energy of 4 MeV
for $\dm$ = $7.5 \times 10^{-5}$eV$^2$ and 3.5 $\times 10^{-5}$ eV$^2$. }
\label{pvsl}
\end{figure}

To trace the reason why the $\theta$ sensitivity is better at $3.5\times10^{-5}$
eV$^2$ in figure \ref{pvsl} we plot the probability vs distance for 
energy fixed at 4 MeV. 
The figure shows that the average distance of $\sim$ 150 km 
of \kl corresponds to 
a maximum in the probability
for $7.2\times 10^{-5}$ eV$^2$ 
while at $3.5 \times 10^{-5}$ eV$^2$
corresponds to a minima. 

The relevant survival probability for \kl is given by the 
vacuum oscillation expression 
\be
P_{ee}=1-\sum_i \sin^22\thsol \sin^2\left(\frac{\dm L_i}{4E}\right)
\label{probkl}
\ee
where $L_i$ stands for the different reactor distances and one needs to do 
an averaging over these. 
Three limits can be distinguished 
\\
$\bullet$  
$\sin^2(\dm L/4E) =  0$ we get a Survival Probability MAXimum 
(SPMAX) \\
$\bullet$  
$\sin^2(\dm L/4E)= 1 $ we get a Survival Probability MINimum 
(SPMIN) \\
$\bullet$  
$\sin^2(\dm L/4E) = 1/2 $ we get averaged oscillation  

In the LMA region the solar $^{8}{B}$ neutrinos undergo adiabatic MSW 
transition and the survival probability can be approximated as 
\be
P_{ee}({\br})\approx \sin^2\thsol
\label{pee8b}
\ee
Whereas the low energy $pp$ neutrinos do not undergo any MSW resonance 
and the survival probability is 
just the averaged oscillation probability in vacuum. 

\begin{figure}[t]
\centerline{\epsfig{file=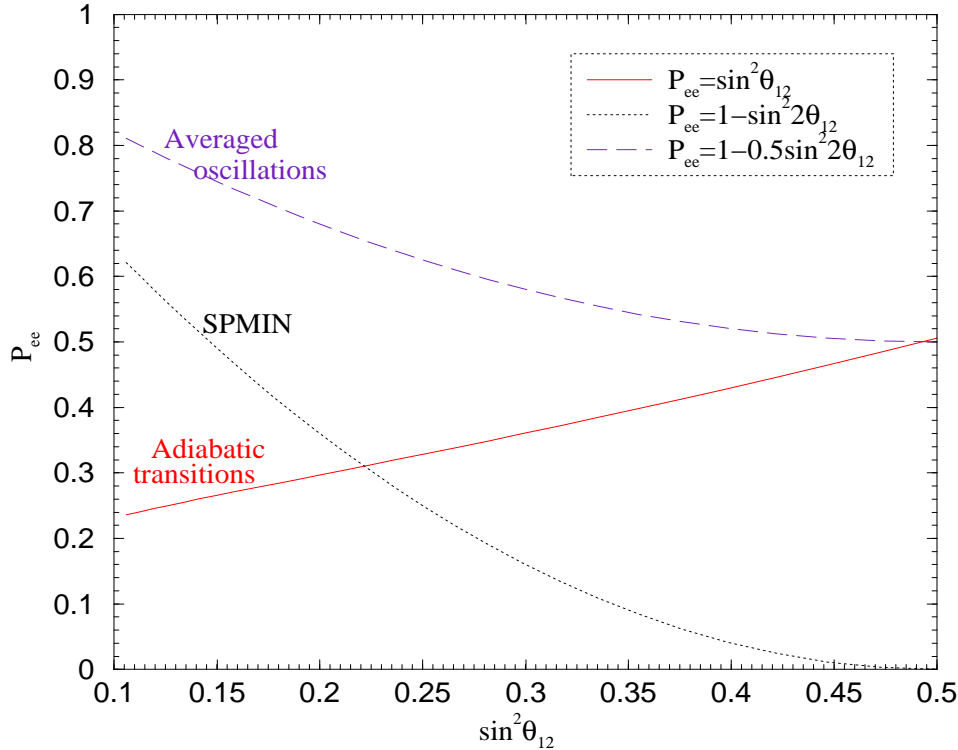,width=5.0in,height=4.in}}
\caption{The survival probability $P_{ee}$ as a function $\sin^2\theta_{12}$.}
\label{sens}
\end{figure}

In figure \ref{sens} we plot the $\theta$ dependence of 
the adiabatic MSW probability as well as the 
probability for the SPMIN and averaged oscillation case \cite{th12}. 
The figure shows that  
for large mixing angles close to maximal,
the adiabatic case has the maximum sensitivity.
For mixing angles 
not too close to maximal ($\sin^2\theta_{\odot} < 0.38$) 
, the $P_{ee}$ for the SPMIN case has the
sharpest dependence on the mixing angle and the \thsol sensitivity
is maximum. Since the 99\% C.L. allowed values
of \thsol is within the range $0.22 < \sin^2\thsol < 0.44$,
SPMIN seems most promising for constraining $\theta_{12}$.
On the other hand at SPMAX the oscillatory term goes to zero and the 
$\theta_{12}$ sensitivity gets smothered.  
Since in the statistically significant region the \kl probability corresponds to an SPMAX for he best-fit value of $7.2\times 10^{-5}$ eV$^2$ the 
$\theta$ sensitivity of \kl is not as good as its $\dm$ sensitivity. 
For this value of \dm   
the SPMIN comes at 70 km. 

\section{A dedicated reactor experiment for $\sss$}

\begin{figure}[t]
\centerline{\epsfig{file=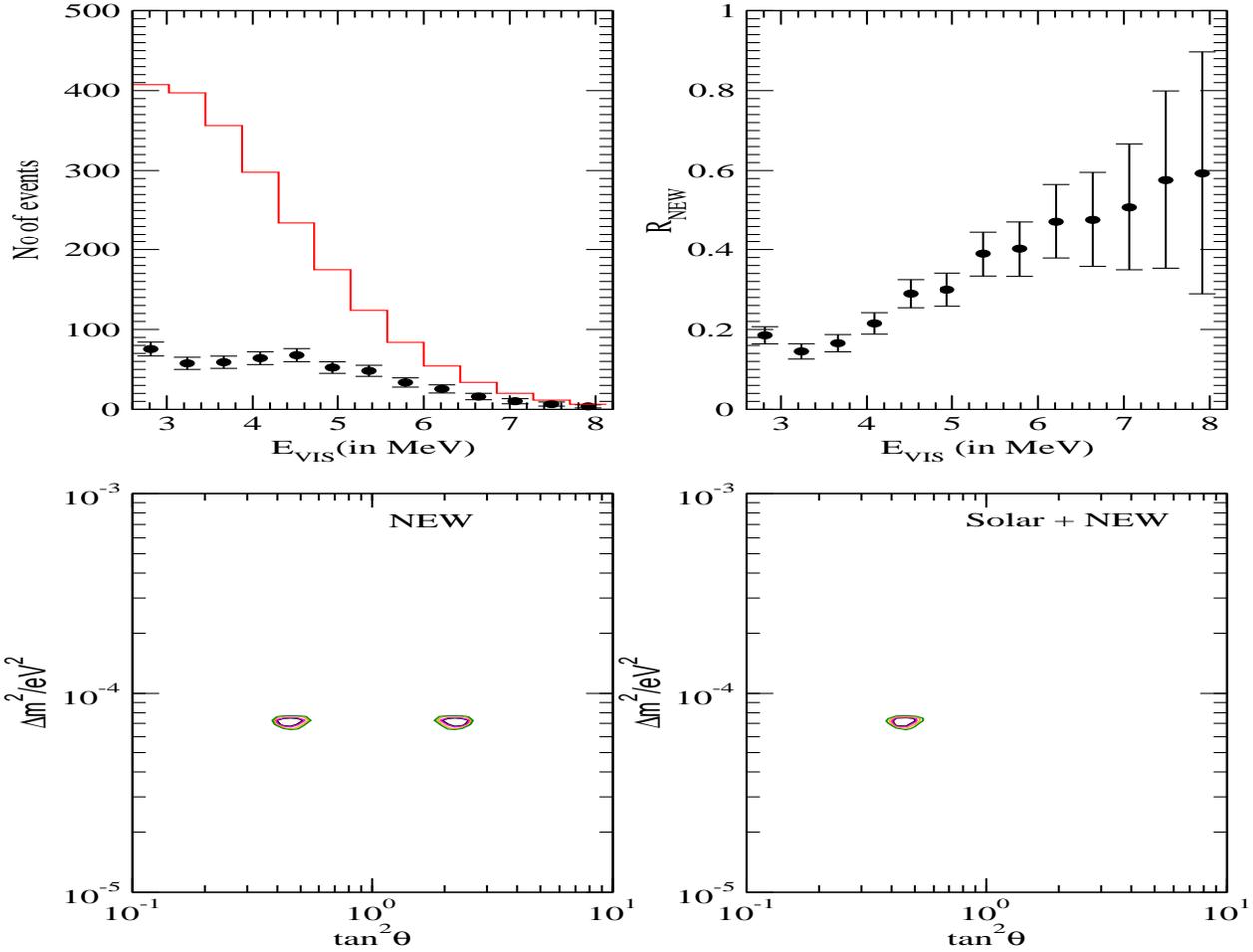,width=7in,height=5.2in}}
\caption{The spectrum and allowed regions for a new reactor experiment with a 
source detector distance of 70 km}
\label{fictnew}
\end{figure}

We show in Figure \ref{fictnew} the constraints on the mass and
mixing parameters obtained using a  "new" dedicated reactor 
experiment
whose baseline is tuned to an oscillation SPMIN \cite{th12}. 
We use the antineutrino flux from a reactor a la
Kashiwazaki nuclear reactor in Japan with a 
power of about 25 GWatt. We assume a 80\%
efficiency for the reactor output and simulate the 3 kTy data
at the low-LMA best-fit for a \kl like detector placed at 70 km from
the reactor source and which has systematic errors of only 2\%.
The top-left panel of the Figure \ref{fictnew}
shows the simulated spectrum data. The histogram shows the expected
spectrum for no oscillations.
$E_{vis}$ is the ``visible'' energy of the scattered electrons.
The top-right panel gives
the ratio of the simulated oscillations to the no oscillation numbers.
The sharp minima around $3-4$ MeV is clearly visible.
The bottom-left panel gives the C.L. allowed areas obtained from
this new reactor experiment data alone. With 3 kTy statistics we find a
marked improvement in the \thsol bound with the 99\% range
$0.39 < \tan^2\thsol < 0.52$ giving a spread of 14\% .
The ``dark side'' solution appearing in the left lower panel 
because of the $\thsol-(\pi/2-\thsol)$
ambiguity in the vacuum  oscillations probability is  
ruled out in the right lower panel by the
solar neutrino data. 
Recently sites of reactor neutrino experiments with a source-reactor distance 
of 70 km has been discussed in \cite{rhone}. 
Also in Japan a new reactor complex SHIKA-2 at $\sim$ 88 km (close to 
SPMIN) will start 
in 2006 
(See however \cite{shika}). 

\section{Other future experiments}

\begin{figure}[t]
\centerline{\psfig{figure=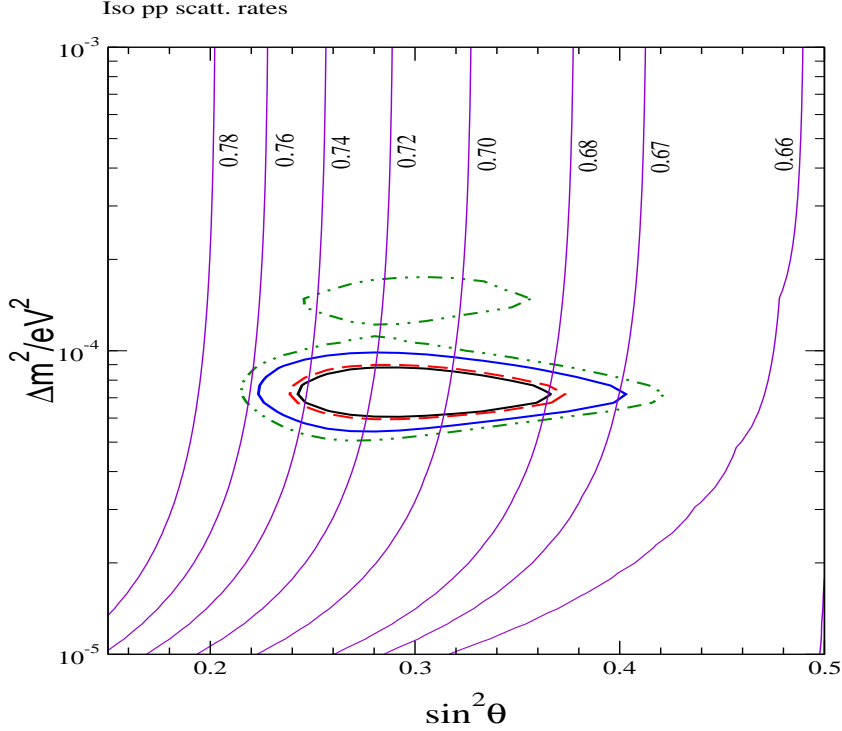,height=5.in,width=5in}}
\vskip -1cm
\caption{The isorates for a  $pp$ scattering experiment.}
\label{lownuiso}
\end{figure}

In Figure  \ref{lownuiso} we show the lines of constant rate/SSM predicted
 in a
{\it generic} LowNu electron scattering experiment sensitive to 
$pp$ neutrinos \cite{lownu}.
At these low energies
 neutrinos the survival  probability  
in the LMA zone is $\approx 1 - \frac{1}{2}\sin^22\theta$ 
and  has almost no sensitivity
to $\Delta m_{21}^2$.
But the $\sss$ sensitivity is quite good and thus these experiments 
may
have a fair chance to pin down the value of the mixing
angle $\theta_{12}$, if they can keep the errors low. 

\section{Conclusions}
The \kl experiment has not only confirmed the LMA solution to the solar
 neutrino problem it has narrowed down the allowed range of $\dmsol$
 considerably 
owing to its sensitivity to the 
spectral distortion driven 
by this parameter.
However the $\thsol$ sensitivity of \kl is not as good. 
Baseline is important to 
identify which parameters would be best determined. We discuss that a 
SPMIN in the vacuum oscillation probability 
is important for the determination of the mixing angle. 
For the current best-fit $\dmsol$ in the low-LMA region SPMIN comes 
at a distance of 70 km \footnote{Note that if high-LMA happens to be the 
true solution contradicting the current trend from the solar neutrino data 
then the SPMIN will correspond to a distance of 20 km \cite{piai}}. 
We propose a dedicated 
70 km baseline reactor experiment to  measure $\thsol$
down to {$\sim 10\%$} accuracy. 
LowNU experiments could be important for
precise determination of {$\thsol$} 
if the experimental errors are low. 
 \\ \\
This talk is based on the work \cite{th12}. The updated 
analysis including the SNO salt results were done in collaboration 
with 
S.T.Petcov and 
D.P. Roy and the authors would like to acknowledge them. 
This work was supported in part
by the Italian MIUR and INFN under 
`Fisica Astroparticellare'' (S.C.).


\end{document}